\documentclass[traditabstract]{aa}
\usepackage{graphicx}
\usepackage{savesym}
\usepackage{amsmath}
\savesymbol{iint}
\usepackage{txfonts}
\restoresymbol{TXF}{iint}
\usepackage{time}
\usepackage{subfigure}
\usepackage{url}

\usepackage[usenames]{color}

\newcommand{\beq}{\begin{equation}}
\newcommand{\eeq}{\end{equation}}
\newcommand{\bea}{\begin{eqnarray}}
\newcommand{\eea}{\end{eqnarray}}

\begin{document}
\frenchspacing
\title{The brightness of magnetic field concentrations in the quiet Sun}
\titlerunning{Brightness of the quiet Sun}

\author{R.S.\ Schnerr\inst{1,2}
   \and H.C.\ Spruit\inst{3}
}
\authorrunning{R.S.\ Schnerr \& H.C.\ Spruit}

\offprints{R.S.\ Schnerr, \email{roald@schnerr.nl}}

\institute{Institute for Solar Physics of the Royal Swedish
  Academy of Sciences, AlbaNova University Center,
  SE-106\,91 Stockholm
  \and
  Department of Astronomy, Stockholm University, AlbaNova University Center,
  SE-106\,91 Stockholm, Sweden
  \and
  Max-Planck-Institut f\"{u}r Astrophysik,
  Karl-Schwarzschild-Str.\ 1,
  D-85748 Garching, Germany
}
\date{\today}

\abstract
{
{In addition to the `facular' brightening of active regions, the quiet Sun also contains a small scale magnetic field with associated brightenings in continuum radiation.}
{We measure this contribution of quiet regions to the Sun's brightness from high spatial resolution ($0\, \farcs 16$-$0\,\farcs 32$) observations of the Swedish 1-m Solar Telescope (SST) and Hinode satellite. The line-of-sight magnetic field and continuum intensity near \ion{Fe}{i} 6302.5 \AA\ are used to quantify the correlation between field strength and brightness. The data show that magnetic flux density contains a significant amount of intrinsically weak fields that contribute little to brightness. We show that with data of high spatial resolution a calibration of magnetic flux density as a proxy for brightness excess is possible.}
{In the SST data, the magnetic brightening of a quiet region with an average (unsigned) flux density of 10 G is about 0.15\%. In the Hinode data, and in SST data reduced to Hinode resolution, the measured brightening is some 40\%  lower. With appropriate correction for resolution, magnetic flux density can be used as a reliable proxy in regions of small scale mixed polarity.}
{The measured brightness effect is larger than the variation of irradiance over a solar cycle. It is not clear, however, if this quiet Sun contribution actually varies significantly.}
\keywords{Sun: surface magnetism -- photosphere -- solar-terrestrial relations}}

\maketitle

%

\section{Introduction}
The brightness of the Sun is known to vary in phase with the sunspot cycle. In terms of the {\em total solar irradiance} measured at the (mean) position of the Earth from the Sun (TSI), it is 0.08\% brighter at sunspot maximum than at minimum spot activity. The implications of solar brightness variations for the Earth's climate are controversial. While a modulation of 0.08\% on the time scale of the 11-yr cycle does not have significant effects, the possible effects of longer-term variations are still being debated either way (cf.\ review in Foukal et al.\ 2006). The brightness of the Sun by direct measurement is unknown on time scales longer than the 30 year record of accurate space-based measurements. Because of the close observed correlation between magnetic activity and TSI, it is possible to make an educated guess of the TSI before 1980 by inference from `proxies': activity indicators like Calcium line emission or the surface magnetic flux, for which longer-term records are available (e.g.\ Lean et al.\ 1992, Chapman et al.\ 1996, Solanki \& Fligge 1998).

The uncertainty in such estimates is that the relation between magnetic fields and  their effect on irradiance is not unique. Large flux concentrations (spots and pores) are dark, the small scale field (plage, network, inner-network) brighter than average, so the mix of small and large has to be known with some accuracy. The brightness of spots is known from observation (see Foukal et al.\ 2006, and references therein); that of small scale magnetic concentrations from theoretical models (Spruit 1976, 1977, Spruit \& Zwaan 1981) and realistic 3-D radiative MHD simulations (Carlsson et al.\ 2004, Keller et al.\ 2004, de Pontieu et al.\ 2006).

The relative amount of small and large concentrations, however, is variable and presently not predictable from theory or observation. This means that contributions to irradiance have to be considered separately for different kinds of surface magnetic fields. In practice, this is done by classifying areas of magnetic activity into `spot', `plage' and `network' components. Plage and network are quantified using a proxy such as brightness in the Calcium H and K lines. They are spread out over such large areas that their contribution to TSI is below the absolute accuracy of the brightness measurements, and their contributions to TSI cannot be quantified directly. Instead, conversion factors of proxy measurement to TSI contribution are introduced to produce a reconstruction of TSI variation from proxy records, and optimized to obtain a best fit with the measured TSI. In this way, 95\% or more of the TSI variability can be reproduced from known manifestations magnetic activity (Fr\"ohlich \& Lean 2004, Wenzler et al.~2006, Ulrich et al.~2010). Within the systematic and statistical accuracy of the data and the proxies used, this value is consistent with 100\%, but its significance is subject to the uncertainty introduced by the use of adjustable proxy coefficients. 

\subsection{Brightness of the quiet Sun}
Particularly uncertain is the contribution from the`quiet network', consisting of small-scale mixed polarity magnetic fields. If this component has a broad distribution over the solar surface, assessing its contribution through a proxy requires measurement in an absolute sense rather than relative to regions assumed to represent the background nonmagnetic Sun level of the proxy used.

Harvey et al.\ (1975), and Ortiz at al.\ (2006) find only low levels of variation of the quiet network. Foukal et al.~(1991) find a variation of 15-20\% between solar maximum and solar and minimum. Withbroe (2009) concludes that the quiet Sun does not contribute more than 20\% of the cycle variation of TSI.  Whether the quiet network, as seen for example in Ca emission varies on time scales longer than the cycle is not known. Lean et al.\ (1992) suggest that it accounts for a brightness excess of 0.15\%, which then might possibly become relevant if the quiet network were to disappear during extended periods of low activity like the Maunder minimum (cf. Foukal et al.~2011). The extended decline of TSI during the last minimum, as compared to previous minima, might be an indication that the magnetic activity of the quiet Sun can vary on time scales longer than the cycle.

These estimates are based on proxies, whose relation with TSI has not been measured in the quiet Sun. Empirical correlations with Ca emission or 10.7 cm radio flux are determined from regions with much higher levels of activity than the quiet Sun. The physics relating these proxies to TSI-relevant brightness is also indirect and theoretically intractable. Because of this lack of a firm physical basis, it is not known if the assumed correlations also extend to quiet regions.

\subsection{Magnetic flux as proxy for brightness}

Magnetic fields measured at the photospheric level are potentially a more tractable proxy, since  most of the contribution to TSI originates at the photosphere. Moreover, small scale magnetic fields such as those in the quiet Sun are now amenable to realistic 3-D radiative magnetohydrodynamic simulations. The excess brightness due to small scale fields (Spruit 1977) can in principle be determined from such simulations (Carlsson et al.\ 2004, Afram et al.\ 2011). Since brightness depends strongly on size of the magnetic concentrations, the main uncertainty in this effort is the size distribution of the magnetic elements (cf.\ Spruit and Zwaan 1981),  which can not yet be determined with existing theory or from MHD simulations.

At high spatial resolution, the observations show more magnetic flux than at the lower resolution of the standard synoptic magnetograms used for long-term monitoring of solar activity. Especially in the quieter parts of the surface, the small scale magnetic field $B$ tends to be of mixed polarity, which averages out in synoptic maps. Use of magnetic flux as proxy thus requires high resolution polarimetric observations. A mean `unsigned'  flux density $\vert \mathrm{B}\vert$ around 10 G appears to be characteristic of areas traditionally called `quiet' (Lites et al.\ 2007, 2008), although the presence of relatively strong horizontal fields seems to be required to explain Hanle measurements (Trujillo Bueno et al.\ 2004).

The brightness effects of interest are quite small, well below the absolute accuracy of the observations, so the analysis hinges on the determination of a reliable reference level.  While the average brightness of the observed area itself does not depend on spatial resolution at all, the reference level does. With decreasing resolution, small scale features of mixed polarity become unrecognizable both in intensity and magnetic flux. Their contribution to intensity ends up in the background level.

The high resolution polarimetric observations required for this analysis are becoming routinely available with Hinode and the Swedish 1-m Solar telescope (SST). Determination of the magnetic brightness excess of typical (very) quiet Sun areas, the accuracy achievable, and complications that have to be accounted for are the subject of the present investigation. One of these complications is the presence of {\em intrinsically weak} magnetic fields, which start to dominate the magnetic signal at the low flux densities of the quiet Sun.

\subsection{Intrinsically weak magnetic fields}

If the magnetic field in quiet regions were made up only of intrinsically strong, kG `flux tubes', brightness excess could be inferred from the same relation between flux and brightness as measured in the kG fields. The `inner network' fields that are an important component in quiet regions do not follow this relation, however. Much of this field must be intrinsically weak (compared with the canonical `kiloGauss fields' in the network boundaries), as shown by their lower center-to-limb variation (Harvey et al.\ 1975) and their ratio of horizontal-to vertical field strengths (Lites et al.\ 2008).  The thermodynamic effects of a magnetic field, and the resulting radiative transfer effects  scale approximately as $B^2$. Per unit of magnetic flux observed, an intrinsic field of 100 G is expected to contribute negligibly to brightness, compared with a kG field.

When using magnetic flux as a proxy, flux has to be corrected for the contribution from such intrinsically weak field fields. This turns out to be possible from the observations themselves. At high spatial resolution a transition from weak to strong fields is clearly identifiable in the average brightness as a function of observed flux density. As described in the following (Sect.~\ref{modeling}), fitting the shape of this `fishhook' dependence (Sect.~\ref{sect:model}) requires the inclusion of a field component that does not contribute to brightness. At lower resolution, such as the MDI observations used by Ortiz et al.\ (2006), the fishhook feature becomes too indistinct to be used in this way.

\subsection{Measuring the brightness contribution of quiet Sun magnetic fields}
Investigations of magnetic brightening of faculae generally use full disk data (e.g.\ Ortiz et al.\ 2006), which have the advantage that the dependence of magnetic brightening on the heliocentric angle can be studied. However, at the 1--2$\arcsec$ resolution of such observations, the solar granulation pattern is barely resolved. Due to the poor spatial resolution it is difficult to differentiate between weak fields covering a significant fraction of a pixel and intrinsically strong fields with a low filling factor. In addition, unresolved mixed polarity fields will tend to average out to a large degree.
To properly assess the magnitude of magnetic brightening in network fields, it is therefore essential to also use observations with a high spatial resolution.

In high resolution images the brightening of individual magnetic elements can be seen directly in the continuum. Counting these and adding up their excess brightness gives an impression of their contribution to the Sun's overall brightness. Measurements of the magnetic flux contributed by such bright points have been reported by S\'anchez Almeida et al.\ (2010).
Magnetic structure smaller than the resolution of the images escapes detection in such a process based on feature identification. In addition, much of the magnetic structure resides in the dark intergranular lanes. Even when brightened relative to their environment, such structures may still be darker than the mean photosphere, and their contribution is also missed in selection based on brightness. For these reasons, measurements based on feature identification would give only a lower limit to the magnetic brightening. 

We improve on these results by combining a statistical approach with high resolution data from the SST and Hinode.

\section{Observations}
\begin{figure}
\includegraphics[width=0.496\linewidth]{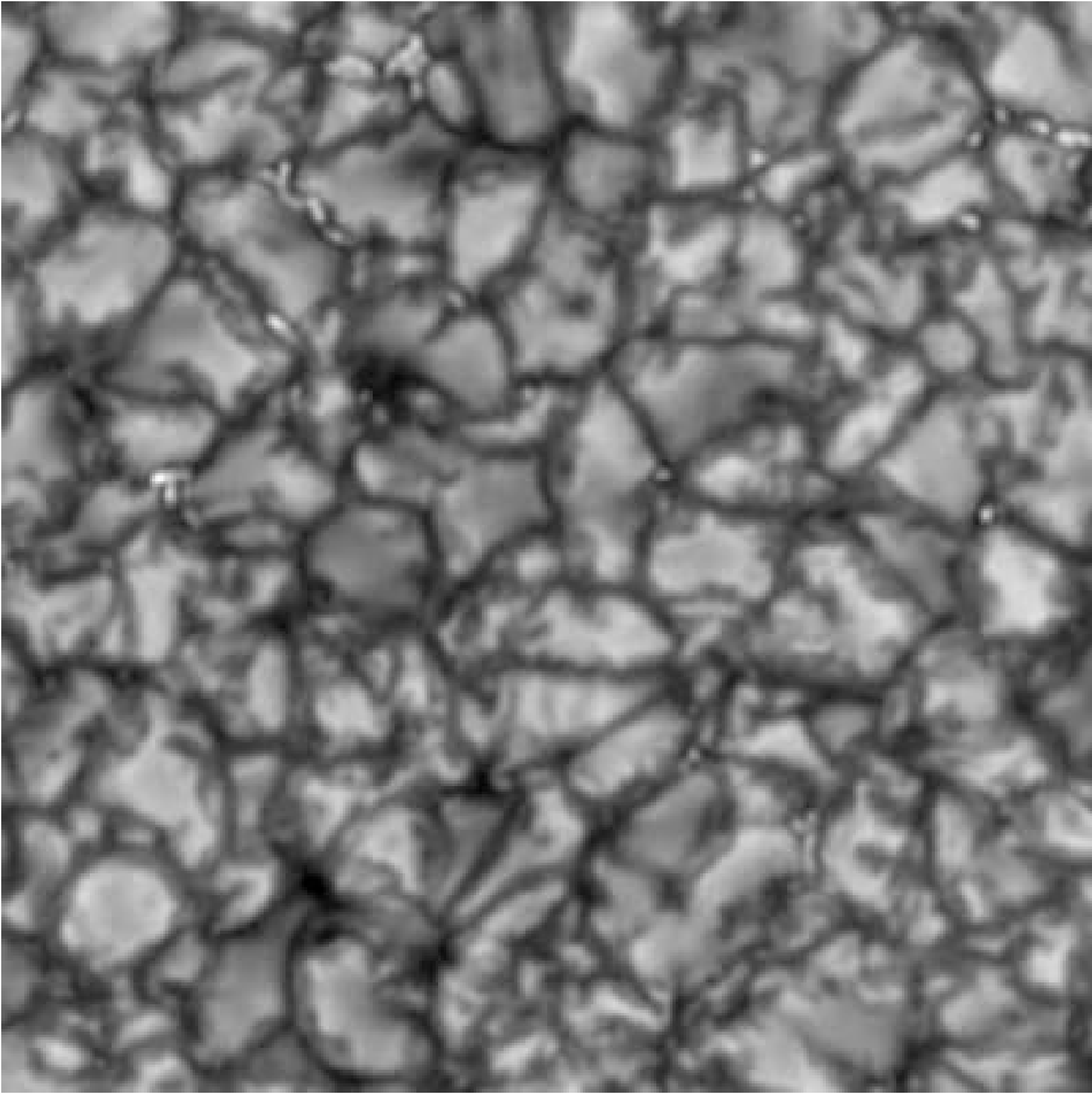}\hfil\includegraphics[width=0.496\linewidth]{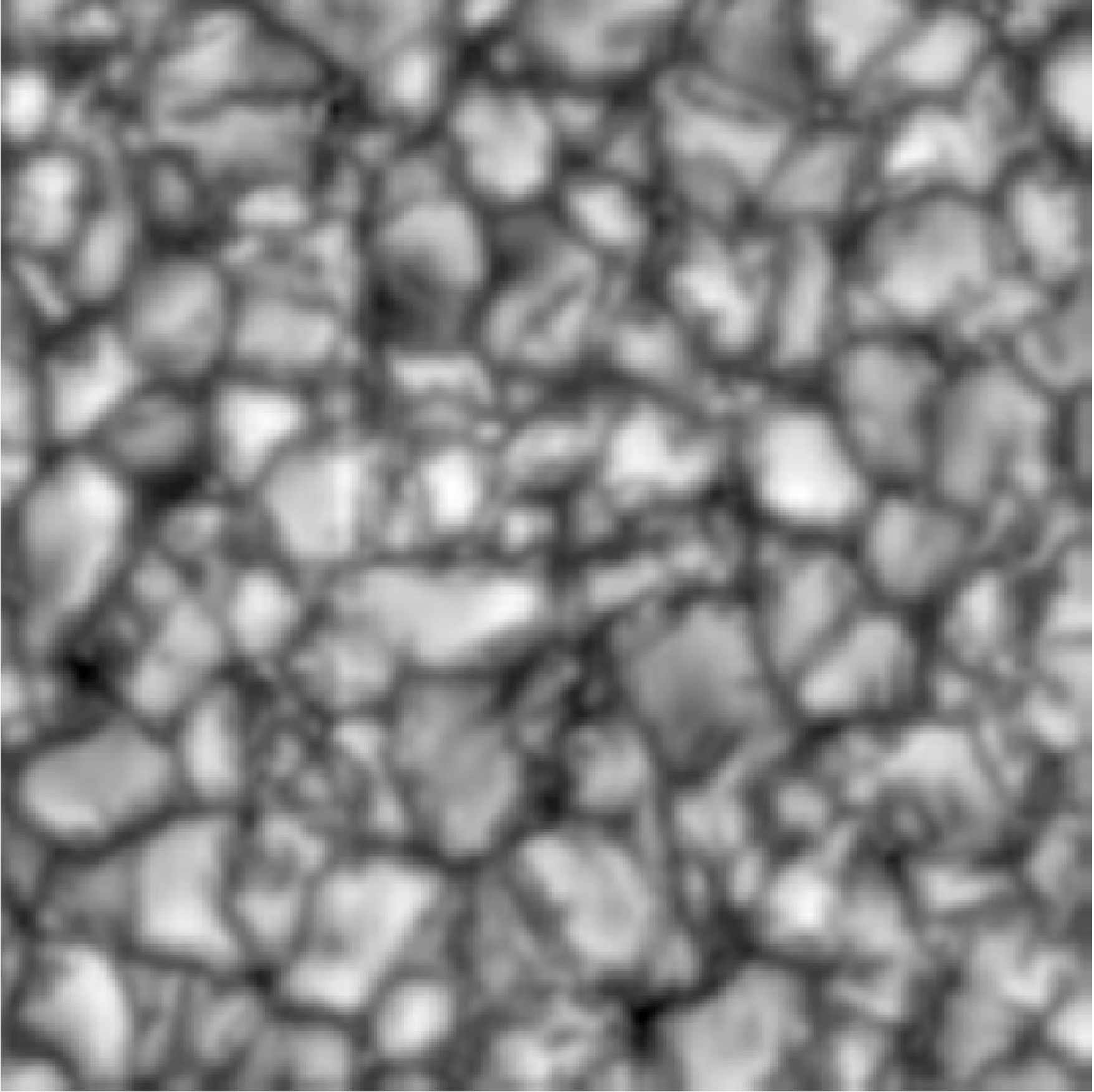}\break
\includegraphics[width=0.496\linewidth]{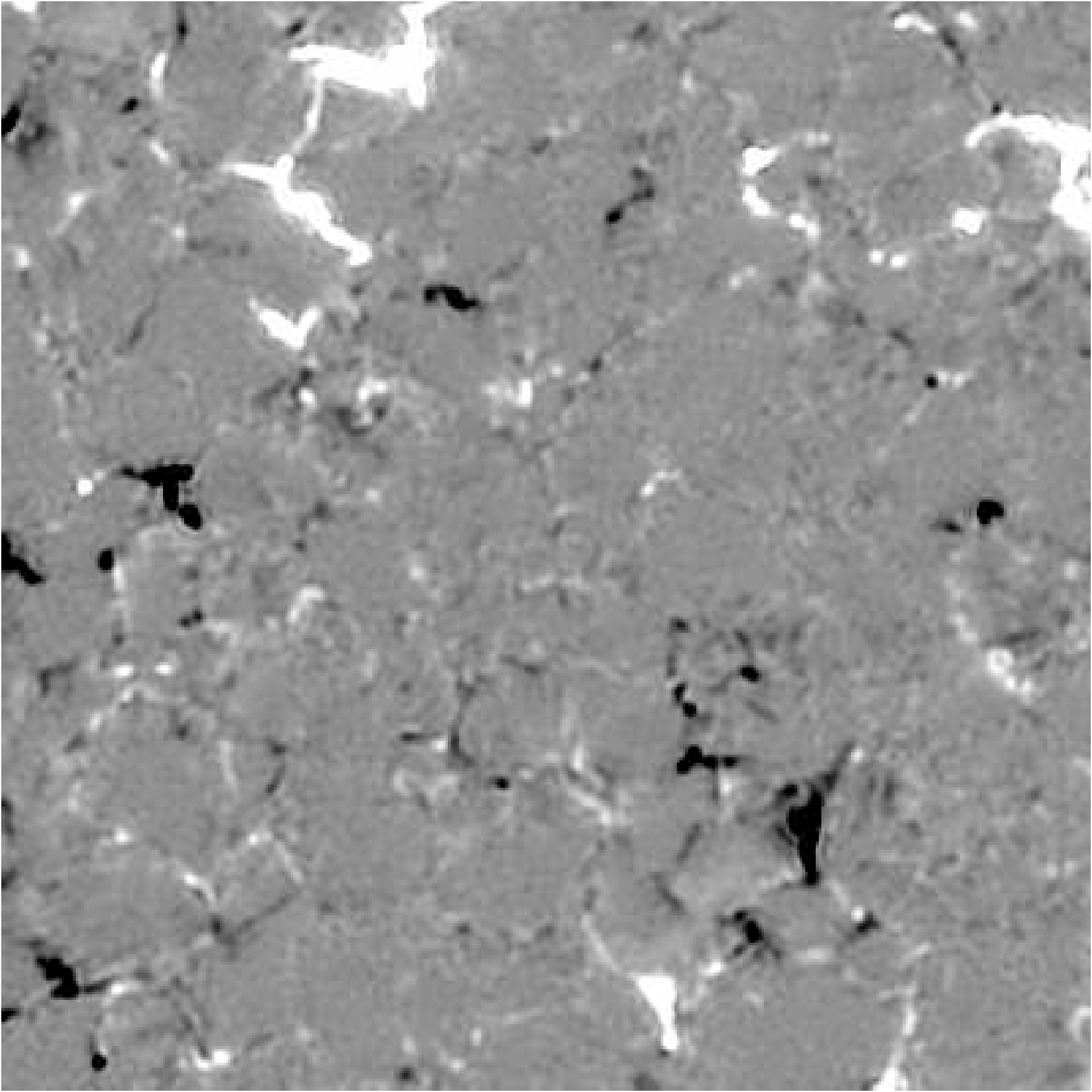}\hfil\includegraphics[width=0.496\linewidth]{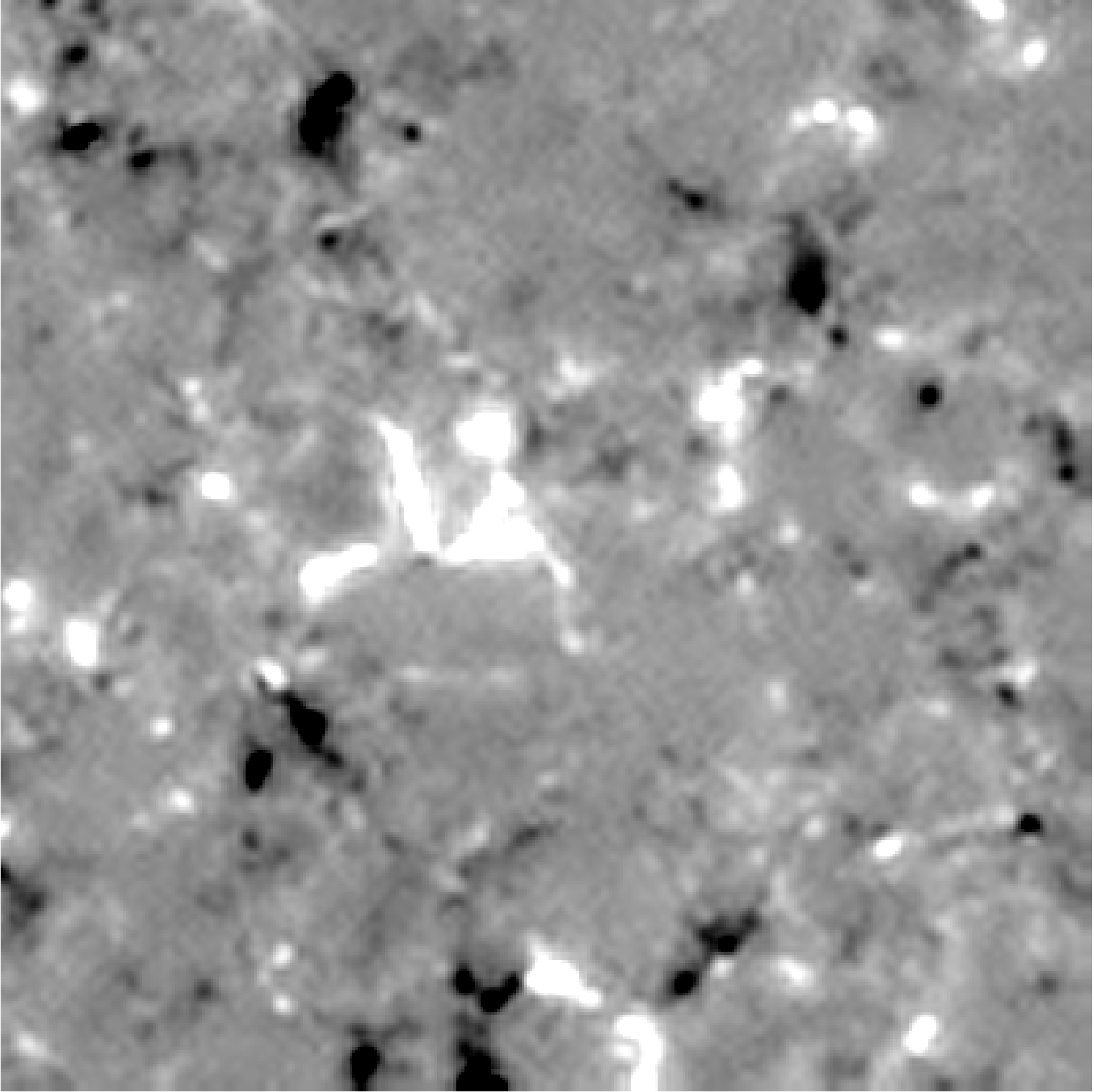}\break
\caption{Quiet regions at disk center as seen in continuum around 630.5 nm \textbf{(top)} and in magnetic flux density \textbf{(bottom)}, showing 20\arcsec x 20\arcsec subfields of the SST (left) and Hinode (right) observations used. Flux density range is from -100 (black) to 100 G (white)}
\label{imco}
\end{figure}

We investigate two disk center quiet Sun fields; one observed with the imaging spectropolarimeter CRISP (Scharmer 2008) on the SST and one with the Spectro-polarimeter (SP) on the Solar Optical Telescope (SOT) of Hinode (see Fig.~\ref{imco}).
The SST observations, obtained on the 23rd of May 2009, cover the 6302.5 \AA\ \ion{Fe}{i} spectral line with 12 equidistant wavelength positions at 48 m\AA\ steps and a continuum point   $\sim$500 m\AA\ to the red side of the line core. The 1k$\times$1k Sarnoff CCD has a pixel scale of $0\,\farcs 0592$, resulting in a total field-of-view of about 60{\arcsec}$\times$60\arcsec. Exposures were recorded with a  frame rate of about 36 Hz, scanning continuously through 4 liquid crystal states and the 13 wavelength positions. At a cadence of about 30 seconds, this resulted in about 1000 frames per observation. The data were restored using multi-object multi-frame blind deconvolution (MOMFBD, van Noort et al.\ 2005) resulting in near-diffraction-limited images.
Restoration and demodulation (for a detailed description of the data reduction procedure, see Schnerr et al.\ 2011) resulted in one Stokes I, Q, U and V image for each of the wavelengths. The noise level in the continuum images of Stokes Q, U and V is 2$\cdot$10$^{-3}$. The rms contrast in the continuum images increased from an average of $\sim$5.3\% before restoration to 8.8\% after restoration. The difference with the predicted contrast from MHD simulations of $\sim$13\% is probably due to straylight and uncorrected high-order modes.

Magnetic field information was obtained from inversions of the data with a parallelized version of the inversion code Nicole (Socas-Navarro et al.\ 2011). Three wavelength points in the telluric line were excluded from this analysis.

The Hinode observations, taken on the 10th of March 2007, cover both the 6301.5 and 6302.5\AA\ \ion{Fe}{i} line, have a pixel scale of $0\, \farcs 16$ and a total (scanned) field-of-view of 164{\arcsec}x328\arcsec. The magnetic field data were taken from the level 2 data products available online\footnote{http://sot.lmsal.com/data/sot/level2dd}. Magnetic field strengths have been converted to fluxes by taking the filling factor into account. This field has already been described by Lites et al.~(2008).

The diffraction limits of SST and Hinode satellite at 6302 \AA\ are $0\, \farcs 16$ and $0\,\farcs 32$, respectively. The average flux density in the Hinode field is 10.8 G, in the SST field 10.1 G. This includes corrections for the measurement noise in the magnetic field which is estimated to be 7 G in the SST data and about half that in the Hinode data. The corrections are small, of order 1 G, because most of the flux appears in fields stronger than the measurement noise.

Since the SST field has a higher resolution than the Hinode data (cf.\ Fig.~\ref{imco}), the flux numbers cannot be compared directly. When convolved to the Hinode resolution, the average unsigned flux density in the SST data drops by 15\%, to 8.5 G  (see Sect.~\ref{rescom} for more on this comparison). The consistency of magnetic flux measurements between Hinode and SST was checked by inverting a subset of the Hinode data with the Nicole inversion code used for the SST data. In this subset we find an average field with a standard deviation of 16$\pm$97 G (Hinode) and 17$\pm$128 G (Nicole) with a correlation coefficient of 0.98, showing that field strengths determined with Nicole tend to be marginally higher.

\section{Modeling magnetic brightening in the quiet Sun}
\label{modeling}

For the analysis of our data we need to use a more sophisticated approach than what is typically done with low resolution data. 
For one, we can not assume that each pixel covers a statistically representative sample of the quiet Sun. The brightness of a pixel will depend not only on the magnetic brightening, but also on whether it is in an intergranular lane or the middle of a granule. This implies that to determine the magnetic brightening we should not compare to the average quiet Sun: a brightened magnetic patch in a dark intergranular lane could still be darker than the average quiet Sun. It is not the average quiet Sun but the brightness that this same patch would have in absence of any magnetic fields, that is relevant. Secondly, even at the high resolution of the SST, flux tubes with high intrinsic field strengths may be unresolved. As discussed above, we need to differentiate between intrinsically strong and weak fields, as the strong fields will have the larger brightness contribution (see Sect.~\ref{sect:weakandstrong}).

We measure the magnetic brightening of the quiet Sun using the information contained in the distribution of the image data in the $I-B$-plane (brightness vs magnetic flux density). This is done via a model that makes use of a number of known properties of the small scale magnetic field, and fitting this to the observed distribution. With a model thus calibrated the magnetic flux density (average field strength in an image pixel) can be used as a proxy for the magnetic brightening in the quiet Sun.

Important information to disentangle the influence of magnetic brightening and the location of the pixel is contained in the run of the average brightness of pixels as a function of their flux density $B$.
We find that at low magnetic flux the contrast first \textit{decreases} with increasing flux before it starts increasing at higher average field strength (see Sect.~\ref{sect:results}), which is not observed at lower resolution (e.g., Ortiz et al.\ 2006). The initial decrease is due to the fact that the magnetic field has a tendency to live in intergranular lanes. This is the case both for the intrinsically weak ($\ll $kG) and strong (`flux tube') magnetic fields. The precise shape of this curve can be used to determine the relative amount of weak and strong field, and the brightness excess of the strong fields as a function of apparent flux density in the image. As described in more detail in Sect.~\ref{sect:model}, this allows the small brightness increase due to magnetic field to be constructed reliably from the data.

\subsection{Weak fields and strong fields}
\label{sect:weakandstrong}
The observed correlation of the Sun's brightness with the small scale magnetic field is understood as a consequence of the local change in radiative energy transfer in the surface layers caused by the magnetic fields (the lower opacity inside the field due to magnetic pressure, Spruit 1976, 1977). The effect has been reproduced in impressive detail in realistic 3-D magnetohydrodynamic simulations (Carlsson et al.\ 2004, Keller et al.\ 2004, see also Steiner 2005, de Pontieu et al.\ 2006). Since the effect scales with the magnetic pressure $B^2/8\pi$, intrinsically strong fields (the kG `flux tubes') have a stronger effect on brightness, per unit of magnetic flux, than intrinsically weak fields. To use magnetic flux as a proxy for brightness, it is thus necessary to separate the contributions of intrinsically weak and strong fields. 

Intrinsically weak fields are seen on the surface in the form of the so-called inner network fields (Livingston \& Harvey 1975, Harvey et al.\ 1975). Their magnetic signal does not show a strong center-to-limb variation, indicating that their orientation is more or less isotropic. This is in contrast with the strong `kG-fields', or network fields, which show a characteristic decline of their Stokes-V signal towards the limb (Martin \& Harvey 1979). This means that strong fields are nearly vertical to the surface, as expected from their magnetic buoyancy (e.g.\ Meyer et al.\ 1979). 

Whereas the strong fields are highly concentrated in the intergranular lanes, the weak fields are somewhat more uniformly distributed. They are advected towards granulation and intergranulation boundaries as expected from a weak magnetic field, but their short life time is compensated by continued emergence inside granular and intergranule cells  (Martin 1988). Phenomena much like those observed have been reproduced in realistic $3{\rm-D}$ MHD simulations by Sch\"ussler \& V\"ogler (2008). These results indicate that local near-surface dynamo action, independent of the solar cycle, may be responsible for the observed weak fields.

\subsection{The model}
\label{sect:model}
From the above it is clear that a model for the distribution of magnetic fields and their brightness contribution needs to take into account the different properties of a weak field and a strong field component. The intrinsically weak component: a) does not contribute to excess brightness, and b) is distributed more uniformly than the strong component. The strong component a) has a brightness excess with respect to its surroundings, and b) is distributed mainly in the intergranular lanes.

We take these properties into account in the following way. The fraction of pixels $f_B$ with a given flux density $B$ is taken from the observations to be fitted. We divide $f_B$ into an intrinsically weak fraction $f_{\rm w}(B)$ and an intrinsically strong fraction $1-f_{\rm w}$. The number of pixels assigned to the weak and strong fractions are thus:
\beq n_{\rm w}(B)=nf_Bf_{\rm w},\qquad n_{\rm s}(B)=nf_B(1-f_{\rm w}),\eeq
where $n$ is the total number of pixels. 
For the dependence of $f_{\rm w}$ on $B$ we take a smooth transition:
\beq f_{\rm w}=e^{-B/B_c}, \eeq
where $B_c$ is one of the model parameters to be fitted to the data. 

The brightness excess of the modeled fields is described by assigning them contrast $q(B_0)$ with respect to their surroundings, where $B_0$ is the intrinsic field strength of the magnetic element (as opposed to the measured average flux density $B$ in a pixel). If $I$, $I_{\rm b}$ are the brightness of the field element and that of the surroundings in which it is embedded,
\beq I=[1+q(B_0)]I_{\rm b}. \eeq
For the weak field component, the model assumption is simply $q_{\rm w}=0$, while the strong field will be given a nonzero brightness contrast. Ideally this should be a function of the size of the magnetic element, since small elements ($\la 0{\farcs}5$) produce a larger brightness excess than larger ones (`pores'). In the larger ones the center becomes dark, as seen in the observations (e.g.\ Spruit and Zwaan 1981) and reproduced in 3-D MHD simulations (Carlsson et al.\ 2004, de Pontieu et al.\ 2006). Contrast $q_w(B_0)$ would thus be a declining function of size, becoming negative in areas with large concentrations of magnetic flux. 

The intrinsic field strength can be retrieved from the data only in sufficiently well resolved structures, however. At the small sizes that have the largest brightness contribution per unit magnetic flux the structures are not resolved at even the best telescope resolution, while their arbitrary location in the image means that most pixels will cover only a part of the structure. At low magnetic flux, we therefore interpret the observed flux density as reflecting {\em filling factor}. The contrast of these pixels is taken proportional to the filling factor or the observed flux density $B$. Together with the observed brightness decline at large flux density, we represent this by the following simple quadratic dependence of the model contrast $q_{\rm s}$ of the strong field component on the observed flux density:
\beq q_{\rm s}(B)=a B(1-\frac{B}{2B_{\rm m}}), \label{intfac}\eeq
where $B_{\rm m}$ is the flux density where brightness contrast peaks, and $a$ an amplitude factor. Both are fitting parameters of the model. At low $B$, the contrast described by Eq.~(\ref{intfac}) is linear in $B$, $q_{\rm s}\approx aB$, and $a$ is the brightness excess `per Gauss of flux density'. 

The  `fishhook' in Fig.\ \ref{haakje} shows an initial steep decline of mean brightness with increasing field flux density $B$, and a subsequent gradual rise. This shape depends critically on the way the magnetic fields are distributed in the granulation pattern. The initial decline shows that the weak field component, from being almost uniformly distributed at very low flux density, favors the darker intergranular lanes  with increasing field strength. Since the weak field component has little contrast relative to its surroundings, magnetic pixels at low flux density are thus darker than average.

The intrinsically strong component has a positive contrast relative to its intergranular surroundings, but at low filling factor the darkness of its surroundings dominates. Though darker than the average photosphere, they still make a positive contribution to the brightness of the area, because they are brighter than the intergranular lane would have been without them.

The model has to fit not just the mean brightness $\bar I (B)$ as a function of flux density, but also the distribution of brightness in the entire $I-B$ plane. For this, probability distributions $p(I_{\rm b},B)$ are needed for the brightness $I_{\rm b}(B)$ of the surroundings of the magnetic elements. We call $p(I_{\rm b},B)$ the `background' brightness distribution; i.e.\ the brightness distribution as it would be in absence of magnetic brightening. At zero magnetic flux this is the distribution of brightness $I_0$ of the non-magnetic Sun, 
\beq p_0\equiv p(I,B=0). \eeq 
We measure it from the image as the brightness distribution of pixels with magnetic flux less than the measurement noise. In units of the average brightness $\langle I_0\rangle$ of the nonmagnetic surface, $I_{\rm b}$ ranges from a minimum $I_1\approx 0.8$ to a maximum $I_2\approx 1.2$. The distribution of brightness in the intergranular lanes where most of the flux resides is not known a priori and must instead be found by fitting to the observed distribution of points in the $I-B$-plane. We find that a suitable starting guess is a `squeezed' version of  $p_0$:
\bea 
p_{\rm ig}(I)=& p_0\left(I_1+(I-I_1)\frac{I_{\rm ig}-I_1}{I_2-I_1}\right) &\quad (I_1<I<I_{\rm ig}),\cr
             =& 0 &\quad(I>I_{\rm ig}).
\eea
where $I_{\rm ig}$ (with value $<I_2$) is a fitting parameter, a measure of the maximum brightness of the intergranular surroundings of the magnetic fields.

For the dependence of $p(I,B)$ on observed flux density $B$, a simple exponential for the transition between $p_0$ and $p_{\rm ig}$ turns out to provide a good fit to the data:
\beq p(I,B)=p_0 e^{-B/B_w}+p_{\rm ig}(1-e^{-B/B_w}),\eeq
were $B_{\rm w}$ is a model parameter determining the width of the transition. 

Not counting quantities like $p_0$ which have been measured from the data itself, the model thus has 5 parameters: $B_{\rm w}, B_{\rm c}, B_{\rm m}, a, I_{\rm ig}$. 
The slope of the initial decline of $I$ with $B$ is controlled mostly by the value of $B_{\rm w}$. $B_{\rm c}$ determines the location of the minimum of the curve, $I_{\rm ig}$ the brightness level of this minimum, $B_{\rm m}$ the location of the maximum and $a$ the brightness level at maximum. Thus 5 is the minimum number of parameters needed for a fit.

\begin{figure}
\includegraphics[width=1.0\linewidth]{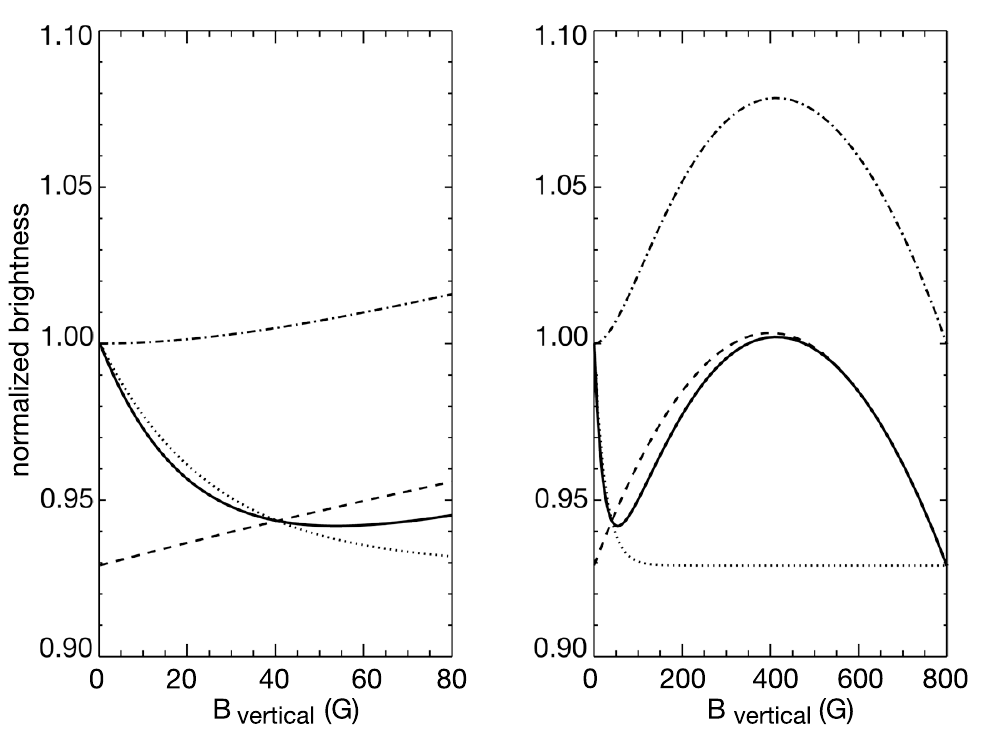}
\caption{Effect of the different ingredients of the model (shown on expanded horizontal scale in the left panel). Solid: model fit of the brightness as function of observed flux density $B$  (the dotted line in Fig.\ \ref{haakje}, Hinode data). Dotted: same if the bright strong field component is omitted from the model. Dashed: same if the weak field component is omitted. Dot-dashed: predicted magnetic brightness if the preference for intergranular lanes were absent.}
\label{contribs}
\end{figure}

The effect of the different ingredients of the model are shown in Fig.\ \ref{contribs}. It shows the model fit (solid line) for the parameter values that fit the Hinode data. The dot-dashed line shows the model prediction if the preference of magnetic fields for intergranular lanes is left out of the model. The dotted line shows how the initial decline of brightness with flux density is due to the increasing tendency of the weak field component to live in the intergranular lanes.
\vskip 2\baselineskip

\section{Results}
\label{sect:results}

\subsection{Fits to the data}
\begin{figure}
\includegraphics[width=0.99\linewidth]{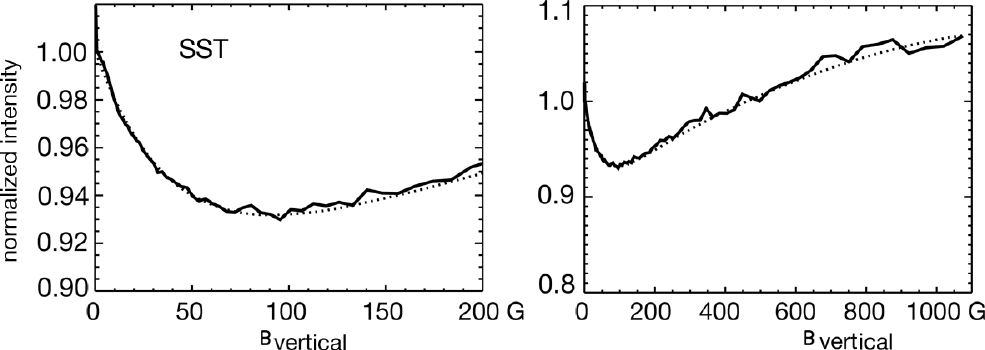}
\includegraphics[width=1.0\linewidth]{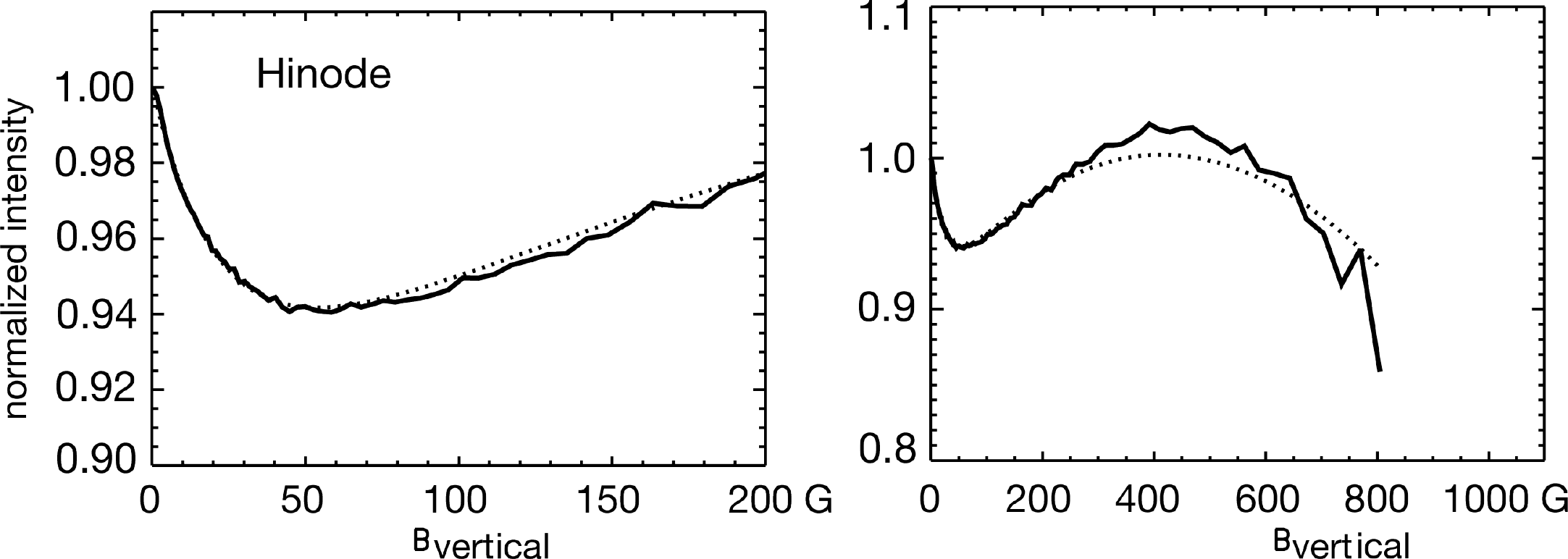}
\caption{Solid: mean brightness as a function of flux density in the SST (top), and the Hinode (bottom) quiet Sun fields (left panel on expanded scale). Dotted: model fits.}
\label{haakje}
\end{figure}

Figure~\ref{haakje} shows the dependence of average brightness $\bar I (B)$ as a function of flux density for the SST and Hinode quiet Sun fields, together with the model fits. For the SST data, the fit yields $B_{\rm w}=80$ G, $B_{\rm c}=100$ G, $B_{\rm m}=1500$ G, $a=4.5\,10^{-4}$ G$^{-1}$, $I_{\rm ig}=1.01$, for the Hinode data $B_{\rm w}=25$ G, $B_{\rm c}=100$ G, $B_{\rm m}=400$ G, $a=4.0\,10^{-4}$ G$^{-1}$, $I_{\rm ig}=1.08$. With these free parameters in the model a good fit can be achieved, so there is no justification for additional parameters.

 The noise in the data can be judged from the point-to-point variation in the curves in Fig.~\ref{haakje}. The uncertainties in the values of the model parameters due to this noise are somewhat correlated. Marginalized over the other parameters, the uncertainty in our main parameter $a$ is of the order 10\% both in the Hinode and the SST field.

At low flux density (left panels) the two fields are similar, but at higher flux density the Hinode field contains relatively more dark structures (pores) than the SST field. The model also provides a good fit to the distribution of points in the $I$-$B$-plane, with these parameter values. At low $B$ the fit is exact (by construction, because of the use of $I_0$ from the data itself). At higher $B$, the model distribution $p(I)$ is a bit narrower than the observations. Experiments with somewhat wider distributions showed that the results depend only marginally on this part of the fitting process, however.

As seen in the lower right panel of Fig.\ \ref{haakje}, the fit for the Hinode field is not very good at flux densities above 300 G. This can be improved by adding an additional parameter in the fitting function (Eq.~\ref{intfac}). The number of pixels at these higher flux densities is small, however, and their contribution to the net brightness effect negligible.

\subsection{Brightness effect}
From the model fits, the excess brightness contribution of the magnetic field can now be computed. Let $n_i(B)$ be the number of pixels in bin $i$ with flux density $B_i$, $n=\Sigma_i n_i$ the total number of pixels, $q(B_i)$ the contrast of the strong field component at this flux density (cf.\ Eq.~(\ref{intfac})), and $f_{\rm s}=1-f_{\rm w}$ the fraction of these pixels that are in the strong field component. The cumulative brightness effect $\delta I (B)$, i.e. the effect contributed by all pixels with flux density less than $B$, is given by
\beq \delta I (B_i)=\frac{1}{n}\Sigma_{j<i} ~n_j f_{\rm s}(B_j) q(B_j). \eeq
The result is seen in Fig.\ \ref{deltI}, which shows the cumulative distribution of excess brightness as a function of flux density. Integrated over all pixels, the effect is  of the order 0.1\%, while pixels with flux density less than 50G still contribute about 0.03\%. The effect is somewhat larger in the SST region than in the Hinode observation.  Since the SST data have a higher spatial resolution than the Hinode data (cf. Fig.\ \ref{imco}), however, the brightness effect of the two cannot be compared directly. 

\subsubsection{Dependence on resolution}
\label{rescom}
\begin{figure}
\includegraphics[width=1.0\linewidth]{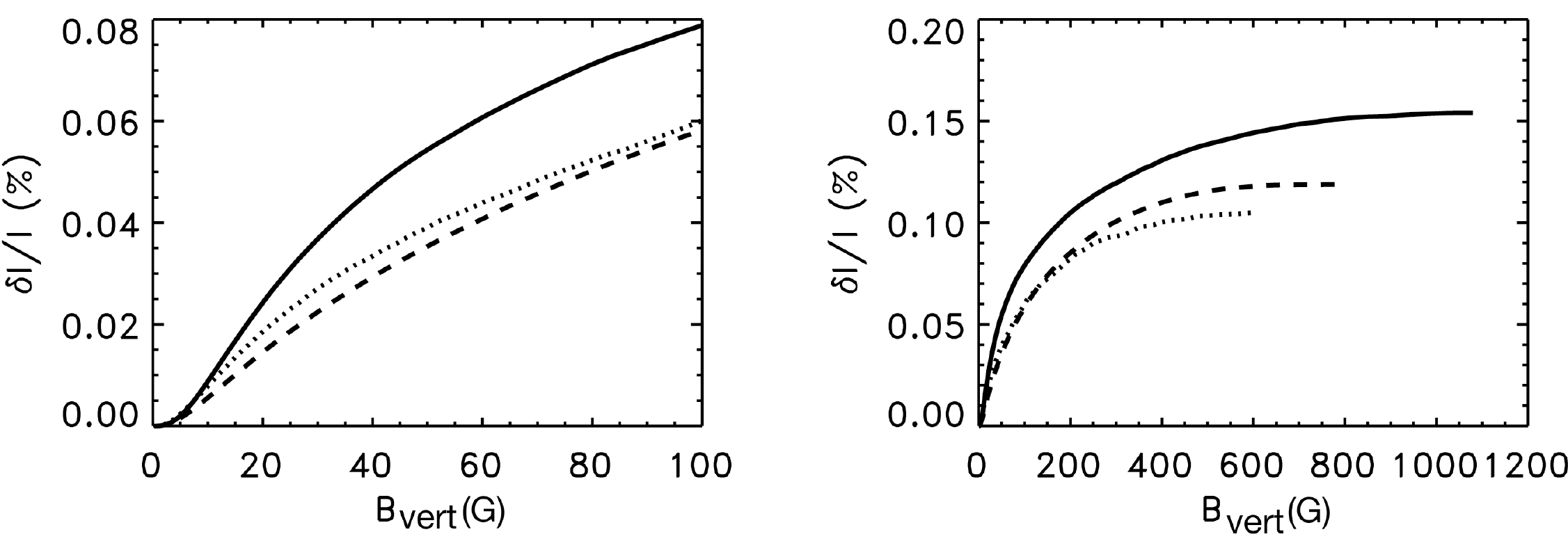}
\caption{Cumulative brightness effect: excess contributed by all pixels with flux density less than $B$, as function of $B$. Solid: SST data, dashed: Hinode data. The dotted line shows the effect of reducing the SST data to the Hinode resolution and pixel scale. The left panel shows a zoom of the right panel.}
\label{deltI}
\end{figure}

To investigate the effect of spatial resolution on the result shown in Fig.~\ref{deltI}, we have repeated the analysis on the SST data after convolving them to a spatial resolution comparable with the Hinode data. The Hinode point spread function has been measured by Mathew et al.\ (2009, see also Wedemeyer-B\"ohm 2008). At  $6300$ {\AA}, its Gaussian core has a width (standard deviation) of $0\farcs 21$ (i.e.\ FWHM of $\approx 0\farcs4$).

Instead of this value, we have convolved our SST data with a somewhat narrower Gaussian, of width $0\,\farcs15$  (roughly 3 SST pixels), and the result rebinned to the Hinode pixel scale of $0\, \farcs16$. The convolution width of $0\,\farcs15$  was chosen such as to yield an rms contrast equal to that of the Hinode image (7.4\%). This is the most relevant measure for comparing our data with the Hinode image, since the brightness contrast of small structures is just the effect we are measuring in this study, and makes the comparison independent of uncertainties in the actual point spread functions of the Hinode and the SST data.

This smearing + rebinning process reduces the average flux density from 10.1 G to 8.5 G. At the same spatial resolution, the SST field is thus actually a bit quieter than the Hinode field (10.8 G). Repeating the analysis on these reduced-resolution data gives the dotted curve in Fig.\ \ref{deltI}. The corresponding values of the fitting parameters are now $B_{\rm w}=35$ G, $B_{\rm c}=100$ G, $B_{\rm m}=600$ G, $a=4.0\,10^{-4}$ G$^{-1}$, $I_{\rm ig}=1.05$. These numbers, as well as the brightness curve itself, are now rather close to those determined from the Hinode data.

The net brightness effect deduced for the area as a whole has decreased significantly by the reduced resolution, from $\delta I/I=1.5\,10^{-3}$ to $0.85\, 10^{-3}$.  Within the systematic uncertainties, the remaining difference in net brightening compared with the Hinode data ($\delta I/I=1.15\,10^{-3}$) is accounted for by the lower average flux density in the SST area.

This comparison suggests that finite resolution still influences the measured brightness effect, even in SST data reconstructed with MOMFBD. A distinction has to be made here between the local effect of a finite PSF and `scattered light' (the distant wings of the PSF). Local smearing dilutes intensity contrast and  polarization in the same way. The correlation between the two as measured here is thus unaffected by such smearing. Dilution over distances larger than the typical separation between patches of opposite polarity adds unpolarized light which reduces all intensity contrasts, but does not change the measured polarization signals. Scattered light thus leads to underestimation of the brightness effect. The rms granulation contrast of 8.8\% in our reconstructed SST data is $\approx 30$\% less than the theoretically predicted value (about 13\% rms), suggesting that scattered light may have reduced the brightness effect as measured with the present analysis by some 30\%.
 
\section{Discussion and conclusions}

Analyzing two regions of quiet Sun with data from the Swedish 1-m Solar Telescope and SOT on the Hinode satellite we find that the mixed polarity magnetic field in the quiet Sun contributes a brightening of about 0.15\% at disk center at $\lambda=6300$ {\AA}. This is more than the variation of total solar irradiance over a solar cycle. 

The method developed here uses the information contained in the dependence of brightness on magnetic flux density as measured from the distribution of image pixels in the $I$ (continuum brightness) vs.\ $B$ (unsigned flux density) plane. It requires the use of some external information, such as observed properties of the `intrinsically weak' field component (discussed below) but does not suffer selection effects, capturing contributions that would be missed by feature identification-based approaches. As a byproduct, it also provides information on amplitude and distribution of the weak field component.

Calibration of magnetic flux density as a proxy for brightness as done in this study is more complex than simple extrapolation of, for example Calcium brightness, but has two major advantages. First, it is close to much better understood physics. Secondly it is becoming accessible to verification with direct numerical simulations.

The observations are sensitive only to the average magnetic flux in a resolution element of the observation, not the intrinsic field strength (at least not at the low flux levels considered), so this distinction cannot be made individually per pixel, but only in a statistical sense.  We have shown how a statistical assessment is possible using the characteristic shape of mean brightness as a function of flux density (Fig.~\ref{haakje}).

An important aspect of this analysis is the distinction between strong, `kilogauss' fields and the intrinsically weak, predominantly horizontal magnetic field component in quiet regions (Martin \& Harvey 1979, Lites et al.\ 2008). Since the thermodynamic effects of a magnetic field scale as $B^2$, the intrinsically weak component is expected to contribute little to changes in brightness, but significantly to the average flux density in the quiet Sun. The model takes this into account by fitting a mixture of the weak and intrinsically strong components,  assuming zero intensity contrast for the weak component.

The results show that spatial resolution has a significant effect on the detectability of magnetic brightening. The  brightening of the SST region analyzed drops by 40\% when the resolution is reduced to that of the Hinode data (see Fig.~\ref{imco}) for the corresponding difference in visual impression of the images). This raises the question how large the uncertainty in the derived brightness is, and which factors are most important for the dependence on resolution. Reducing  resolution does not change the brightness of the image, but affects the background level. In regions of unresolved mixed polarities the magnetic flux is underestimated and their excess  brightness contribution instead gets lumped into the background level.  This is probably the main reason for the decrease of the brightness effect at lower resolution.

The brightness effect measured is monochromatic at $6302$ {\AA}; the contributions at other wavelengths would need to be considered as well for a more quantitative estimate of the effect on solar irradiance (TSI). 

The data show that in regions as quiet as those studied here there is a significant contribution to the magnetic flux density from intrinsically weak fields that do not contribute to brightness. Quantification and correction for this effect is needed when using the flux density as a proxy for brightness.

\subsection{Contribution to TSI variation?}
Whether the brightness effect found here is of importance for solar irradiance depends on the degree to which it varies, especially on the time scale of the solar cycle or longer. It is not obvious that the increase of 0.15\% over a hypothetical field-free Sun, though of the same magnitude as the variation of TSI over a cycle, has much practical effect since the Sun is not observed to be field-free even at minimum activity. 

In the numerical simulations of Sch\"ussler \& V\"ogler (2008), Pietarila Graham et al.\ (2009), the advection and subsequent compression of weak fields into the intergranular lanes generated field strength up to kG values. This raises the possibility that a part of the intrinsically strong field identified in our analysis actually represents this weak field dynamo process rather than solar-cycle related mixed polarity fields. The brightness-relevant fields in quiet Sun measured here would then be stable in time, hence irrelevant for TSI. On the other hand, it is known from synoptic data that simple decay of active regions by dispersal of its magnetic flux contributes directly to the quiet Sun as seen for example in the Calcium network. In addition, a large amount of short-lived magnetic flux appears in the form of ephemeral active regions (Harvey et al.\ 1975). There must thus also be a contribution to the quiet Sun magnetic field that is related to the solar cycle. 

\medskip
Not addressed in all of the above is the possibility of secondary effects of magnetic fields on brightness: effects that would be due to the presence of the field, but not strictly cospatial with it. Potentially most worrying of these is the effect of strong magnetic flux bundles on the pattern of convection around them. Minor effects on convective transport efficiency due to embedded flux bundles, too low to be measured directly on the Sun or in simulations, might still be large enough to have a noticeable net brightness effect. 

Another such effect are the narrow dark rims seen at high resolution around magnetic brightenings. These are understood as consequence of radiation leaking into the magnetic element from the sides; they compensate the brightening to some extent (Spruit 1977). Through this effect, measurement  at the highest spatial resolution may actually {\em overestimate} the magnetic brightening somewhat. `Proximity effects' like this are taken into account implicitly in proxy data calibrated against observed TSI variation, but a more direct assessment of their importance would be preferable.

\acknowledgements{We thank the referee for critical comments, which have led to significant rewriting of the text. We thank Peter Foukal for extensive discussion of the results and their interpretation. R.S.\ would like to thank C.E.\ Fischer. The Swedish 1-m Solar Telescope is operated on the island of La Palma by the Institute for Solar Physics of the Royal Swedish Academy of Sciences in the Spanish Observatorio del Roque de los Muchachos of the Instituto de Astrof\'{\i}sica de Canarias. Hinode is a Japanese mission developed and launched by ISAS/JAXA, with NAOJ as domestic partner and NASA and STFC (UK) as international partners. It is operated by these agencies in co-operation with ESA and NSC (Norway).}

\bibliographystyle{../aa-package/bibtex/aa}

\end{document}